# COLLECTIVE DYNAMICS OF HIERARCHICAL NETWORKS: LESSONS FROM DISASTER RESPONSE EXERCISES

***Liaquat Hossain (Corresponding Author)***
Professor, Information Management
Division of Information and Technology Studies
The University of Hong Kong
lhossain@hku.hk

Honorary Professor, Complex Systems
School of Civil Engineering
Faculty of Engineering and IT
The University of Sydney, Australia
Liaquat.hossain@sydney.edu.au

***Rolf T Wigand***
Maulden-Entergy Chair & Distinguished Professor
Departments of Information Science & Management
UALR, 548 EIT Building
2801 South University Avenue
Little Rock, AR 72204-1099, USA

**ABSTRACT**
In an increasingly complex, mobile and interconnected world, we face growing threats of disasters, whether by chance or deliberately. Disruption of coordinated response and recovery efforts due to organizational, technical, procedural, random or deliberate attack could result in the risk of massive loss of life. This requires urgent action to explore the development of optimal information-sharing environments for promoting collective disaster response and preparedness using multijurisdictional hierarchical networks. Innovative approaches to information flow modeling and analysis for dealing with challenges of coordinating across multi layered agency structures as well as development of early warnings through social systems using social media analytics may be pivotal to timely responses to dealing with large scale disasters where response strategies need to be viewed as a shared responsibility. How do facilitate the development of collective disaster response in a multijurisdictional setting? How do we develop and test the level and effectiveness of shared multijurisdictional hierarchical networks for improved preparedness and response? What is the role of multi layered training and exercises in building the shared learning space for collective disaster preparedness and response? The aim of this is therefore to determine factors that may be responsible for affecting disaster response. It is shown here that changes to the interconnectedness of nodes in the network may have implications on the potential to preparedness and response. In this study, theory behind social network analysis is applied to a large-scale inter-organisational Disaster Response Network (DRN) for exploring correlation between network interconnectedness and response. We discover that the leadership and involvement displayed by organisations in multijurisdictional emergency response is not equal, and hypothesise the existence of a loose tiered structure that guides how interconnected an organisation should be. A model is presented as a theoretical means to confront the issues of disaster response. To test our hypotheses, we investigate survey data from state law enforcement, state emergency services and local law enforcement by performing agency-based (macro) and cross-agency (micro) analysis to identify attributes of each network and response. Results suggest that there is a positive correlation between network connectedness and potential to effective response as well as the concept of tiers within DRN may exist which can be characterized by the sub-network that an organisation associates with.

**DISASTER RESPONSE NETWORK (DRN)**

The aim of the Disaster Response Network or DRN is to respond to extreme events as quickly and efficiently as possible in order to return society to a "business as usual" state, and thus restoring social confidence and economic stability (Waugh, 2003). *Consequence Management* can be triggered by a disaster serving as motivation for establishing DRN with the condition to save lives of the victims involved. Consequence management requires the mobilization of a complex network of organizations designed to be able to form rapidly to coordinate a multifaceted disaster response and then quickly dissolve once the incident has been controlled (Kapucu, 2003; Kapucu, 2005). The effectiveness of consequence management may be defined through quantifiable measures such as the number of lives lost, property damage, or perhaps overall time till recovery, which is referred as society's "resilience" to extreme events, and are the gauge for assessing a society's ability to cope (Wildawsky, 1971; cited in Kapucu, 2005).

Van Scholten et al. (2005), Comfort and Kapucu (2003) and Waugh (2003) study on the state of American DRN identified several existing flaws that require further development for effective disaster relief and response. The specific challenges of DRN spur from the networks need for cooperation and coordination within the complex network of interdependent organizations. Van Scholten et al. (2005) describe common horizontal cooperation issues such as police, fire and medical crews having problems when interaction is necessary in the face of responding to interdependent tasks that are outside the scope of any one organization, especially when coordination is necessary in the face of a highly stressful and turbulent environment such as a disaster. Waugh (2006) further suggests that cooperation flaws are also apparent within the communication and coordination of organizational actors across vertical levels of leadership, such as federal government agencies working with nonprofit organizations. The ability to coordinate between sectors such as with unaffiliated volunteers can potentially create a significant challenge and put strain on the coordination of the network. This is supported by literature revealing that some organizational actors within the network are reluctant to rely on other sectors in times of disaster and crisis (Kapucu, 2005). Moreover, it is especially directed towards nonprofit agencies, which according to Waugh (2006; cited in Kapucu, 2005), maintains an assertion from other sectors as being poorly skilled, lacking resources, and having the potential to inhibit the response by placing themselves in danger or obstructing professionals in the response effort.

The ability for DRN to preserve sufficient information flow in the network rests in its ability to maintain a structured and stable distributed network where all avenues of communication remain open. Kapucu's (2005) study into the 9/11 documents the challenges of such tasks when faced with a disaster event of significant magnitude. The problems associated with maintaining information flow produced a direct effect on the ability for leading organizations in the network to make informed decisions based on whole information, and other organizations to work together to carry out the directions (Van Scholten et al., 2005). Van Scholten et al. (2005) argue that decisions must be made in extreme events regardless of circumstance, and results of an impaired communication network has the ability for poor decisions to be made because of incomplete or even wrong information,. This, moreover, could have a significant impact on the efficiency of a response effort.

The lessons learned from current literature identifies the need for communication across vertical sector-based boundaries and horizontal same-sector organizations to minimize the prevalent coordination gap apparent within the current workings of DRNs (Van Scholten et al., 2005; Kapucu, 2005; Kettl, 2006). Studies into the events of Hurricane Katrina and September 11 illustrate that under both naturally occurring and man-made disasters, the establishment of DRN alone (Stanley, 2006; Kapucu, 2005) is insufficient to create a resilient society to overcome extreme disaster events in a timely and efficient manner, but it is the development of coordination within the network that facilitates this process. The network structures outlined in the FRP (see Figure 2. Kapucu, 2005; Department of Homeland Security, 2006) illustrates that in the event of a crisis, the emergency agencies involved in consequence management are to quickly unite and form a distributed network where all agencies are central to the flow of information and are to coordinate themselves to respond to any interconnected set of problems that may present itself from the situation. The reality of this plan, however, was unable to reproduce the sophisticated, yet simplistic network as intended. The research conducted by Kapucu (2005) looked at the situation reports of the September 11 crisis, which represents an 'actual' network as it stood during the event (Figure 1, Kapucu, 2005). The network graph displays a contrast between the planned and the actual network with a significant lack of interconnectedness and communication flowing between agencies. Granovetter (1983) suggests the implications of a reduced number of connections (or network ties) can potentially lead to a reduction in coordination due to actor segregation which limits the flow of information. This is especially important in DRN to create a sense of community and share knowledge in order to overcome problems that require an interdependent multi-agency response. Kapucu's (2005) study into DRNs uses an exploratory model to assess interconnectedness of organizational actors within a network during a crisis. The ability of Kapucu (2005) to look at actual rather than perceived data and interpret how the network functions during an actual incident is very useful for investigating how planned actions present themselves in live situations. The limitation of Kapucu’s (2005) study, however, is that it does not allow for an assessment of coordination within the network as an outcome of interconnectedness derived from network planning and optimization. It imposes challenges in judging the success of an organization in performing its role in the network during a crisis.

Figure 1. Graphical Representation of Organization Network – FEMA Situation (in Kapucu, 2005)

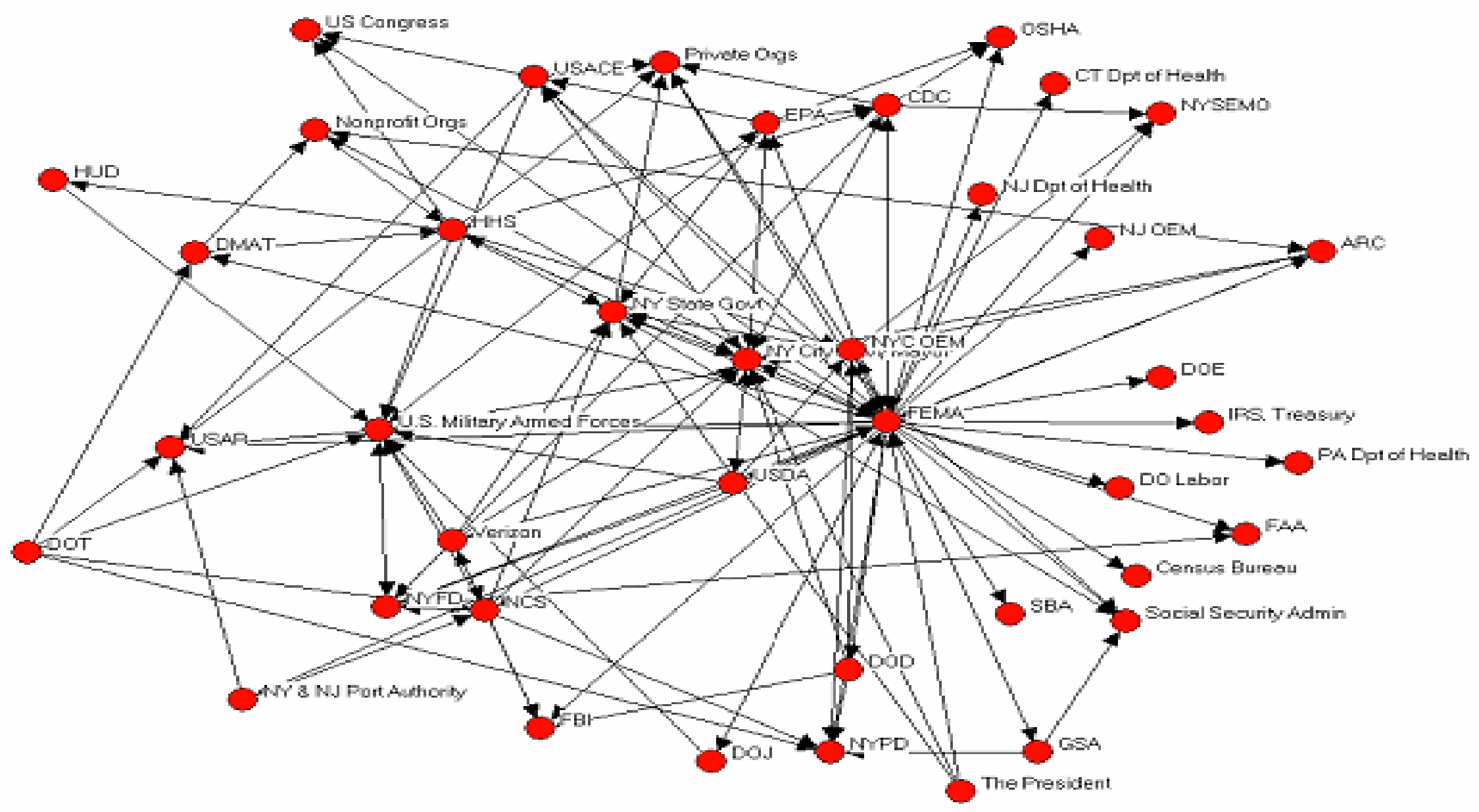

## MODEL FOR DISASTER RESPONSE NETWORK

The model is constructed with a view to assess the current state of preparedness and response as a product of the attributes of the network. The framework for the model is intended for DRN assessment during a non-crisis period in order to optimize network performance by creating a heightened state of preparedness. The model may be applicable to other networks that is distributed in nature and requires response to produce better performance. The model, as illustrated in Figure 2, depicts a framework for investigating disaster preparedness based on network connectedness (evaluated through social network analysis). There is a single moderating variable defined as “Tiered Organization” which places an organization in one of three DRN tiers, which then forms the basis for assessing whether the resulting level of network involvement and thus potential to coordinate the network is adequate for a given agency. The connectedness of an agency within the network is measured by the three independent social networking variables which together produce an organizational actor’s assessment of network involvement. The three dependent variables define the characteristics of an organization’s current state of coordination and coordination potential in an emergency. The aim of applying the framework to DRNs is to empirically investigate the relationship between the network itself and the potential for response. The driving theory for constructing the model is based on the view that enhancing network performance correlates to increasing the capacity for coordination to occur. As a result of increasing network performance, the implied coordination gap present in emergency networks (Kapucu, 2005; Waugh, 2006; Kettl, 2006; Rathnam et al., 1995) may be reduced.

Figure 2. A Model for Assessing Disaster Preparedness and Response Networks

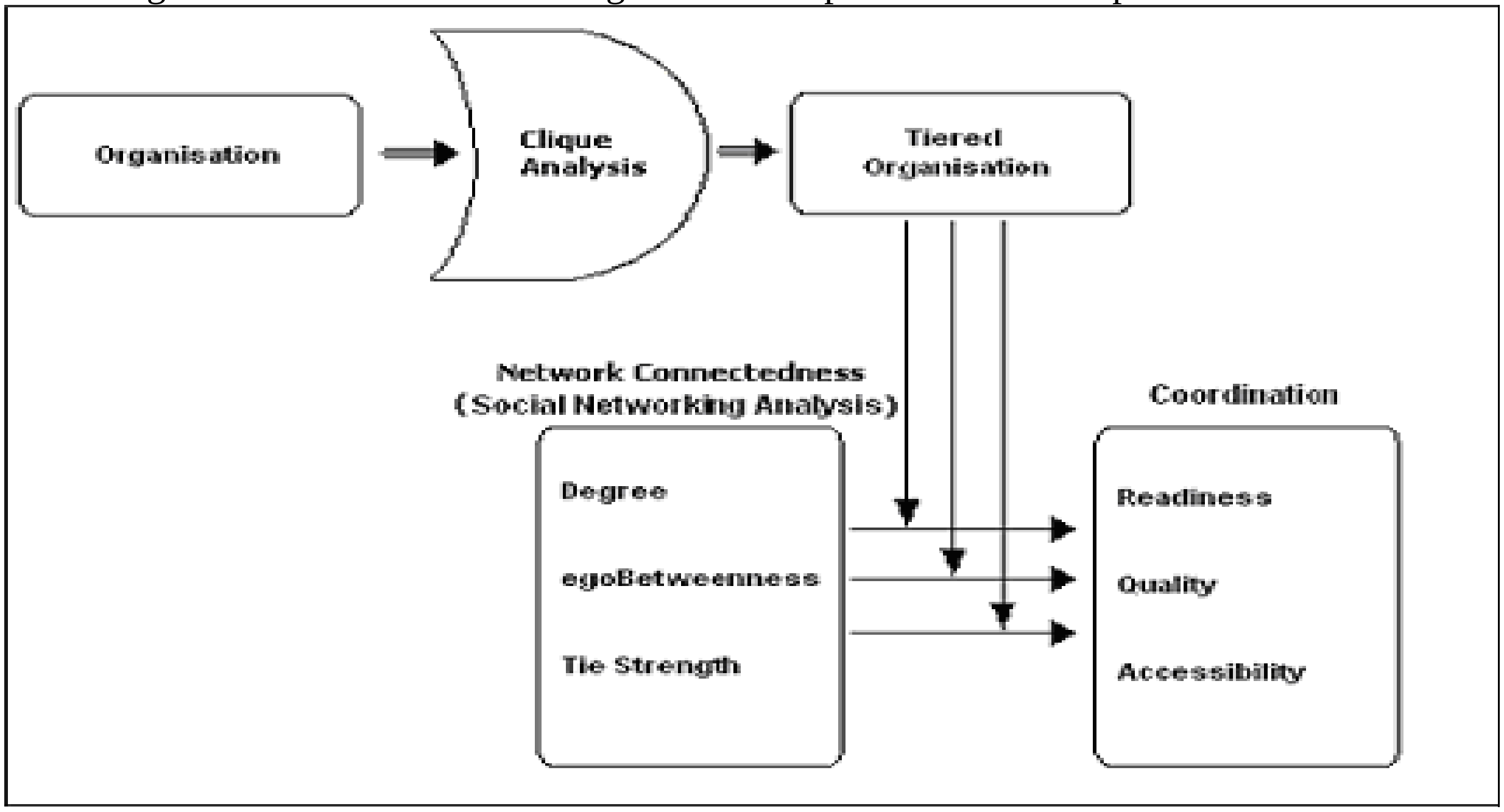

An organization in the model represents any agency for any business sector that is somehow involved in the disaster network, in which each and every organization within DRN can be assessed to determine their current state of preparedness and therefore, required to alter their network connectedness level based on where they should be operating. A clique analysis defined as a sub-set of actors within a network that are more closely tied to create a subgroup is then carried out to assign tier allocation (Hanneman et al., 2005), which assist in assessing the cluster an organization belongs to. According to the National Response Planning or NRP (Department of Homeland Security, 2006) each agency involved in consequence management is responsible for taking on certain leadership roles within the DRN. The NRP implies a loosely based leadership structure that is assumed to represent three tiers of responsibilities whereby federal organizations represent the first tier, followed by state and local agencies, and then all other sectors and organizations. The clique analysis of an organization is hypothesized to provide evidence of tier placement based on an actor's subgroup association. The reasoning behind this theory is that if an organization communicates with significantly more state and local authorities, it is likely that the agency being assessed also belongs to that particular tier. This leads to the first hypothesis that:

(H1) – *A prediction of what tier an organization belongs to can be made by analyzing an organization for its subgroup*.

A tier level determines an organization's expected level of network connectedness. We use tiered organization as moderating variable in this study. A tiered organization represents an agency that has been placed in a particular tier group that determines its expected level of network involvement. Based on the literature from the NRP (Department of Homeland Security, 2006) an association with a particular tier retains expectations of network leadership and coordination facilitation. By this reasoning, an assessment of preparedness is imbalanced across all organizations as performance expectations are not equal and therefore, a weighted score based on tier positioning is required. This leads to the second hypothesis that:

(H2*) The level of network involvement expected from an organization is mediated by the tier they fall into*.

We then use network connectedness as an independent variable in our model. A degree centrality analysis is carried out to determine aspects of an ego's network**.** Degree as an independent variable is used to measure connectedness, in which we defined the number of relationships (also known as arcs or ties) that a particular node (actor) is connected to. Marsden (2002) mentioned that the identical degree centrality measure may be used for an egocentric analysis as the principal for defining relationships remains the same. EgoBetweenness analysis is then carried out to determine aspects of an ego's network, in which the measure of betweenness characterizes the extent to which a node lies in between other nodes in the network, or the extent to which a node falls on the shortest path between pairs of other nodes (Chung, Hossain and Davis, 2005; Freeman, 1977). Lastly, a tie strength analysis is carried out to determine aspects of an ego's network.

An analysis of readiness is carried out to determine aspects of an actor's current state of coordination**.** We use readiness as dependent variable in our study, which defines readiness as an organization's perceived ability to react to a crisis event should the need arise at any given moment. The variable is based on the literature by Kapucu (2005) where research is presented providing evidence that simple planning does not amount to a state of readiness when the emergency need arises. Waugh (2006) and Kettl's (2006) account of the Emergency Response Network (ERN) in hindsight of emergency disasters generate discussion on particular agencies' roles and the need for a faster response in order to create resilience. This concept has been recognized and characterized here as actors readiness.

An analysis of the quality of information received is carried out to determine aspects of an actor's current state of coordination**.** In our framework, we define quality (dependent variable) in terms of the quality of information that is received by the agency under assessment in order to examine what coordination benefits they can provide. Scholten et al. (2005) mention the coordination benefits of improved quality of information disseminating through the network as an improved ability for collaboration and decision-making strategies. We further perform an analysis of the accessibility (dependent variable) of information. This is carried out to determine aspects of an actor's current state of coordination. Accessibility of information refers to an organization's capacity to retrieve information from a multitude of sources. The importance of being granted access to information in DRN is that regardless of information, decisions must be made about how the network will be coordinated to respond to a crisis. By having access to different sources, organizations are able to group disparate pieces of information to develop a more whole understanding of the problem and be in a more informed position to make urgent decisions (Scholten et al., 2005). By investigating the three measures of social networking theory that combine to represent an organizational assessment of network connectedness, it is theorized that the involvement of an organization in the DRN has significant implications on the coordination performance it is able to achieve. The hypothesis proposed as a result of this concept is that:

(H3) *There is a significant relationship between network involvement and coordination where an increase in network connectedness produces an increase in organizational coordination within a given threshold.*

**DISASTER RESPONSE NETWORK DATASET**

The dataset entitled “Domestic Terrorism: Assessment of State and Local Preparedness in the United States, 1992” was found at the Inter-University Consortium for Political and Social Research (ICPSR) website: (http://www.icpsr.umich.edu/cocoon/ICPSR/STUDY/06566.xml, last accessed on *15 April 2008*. The study was developed with the purpose to “analyze states' and municipalities' terrorism preparedness as a means of providing law enforcement with information about the prevention and control of terrorist activities in the United States” (Riley and Hoffman, 1995). The study was funded by the United States Department of Justice and the National Institute of Justice. Research investigation was carried out by Kevin Jack Riley and Bruce Hoffman of the RAND Corporation. The research agenda was to conduct an assessment of how state and local law enforcement perceived the threat of terrorism under the federal level of government. The framework for data collection involved sending each selected agency a package, which included the survey instrument, a request letter of participation, a confidentiality agreement, and a brief overview of the RAND Corporation (Riley and Hoffman, 1995). The procedure after the initial invitation was to follow up with a second letter after a ten day period as a reminder notice. Should the study not be filled in within three weeks, another package containing all documents found in the original invitation were resent, and was subsequently followed two weeks later by a phone call to the agencies who had not yet responded. It is noted that if the survey still was not filled in and returned twenty days after the final telephone call, another jurisdiction would be contacted to replace the non-respondent agency.

The sampling technique used to invite participants involved a two-part methodology for local law enforcement agencies. The first stage of sampling required the selection of twelve counties in each census district using a population-based method. Three counties were selected based on the 1990 U.S Census estimates that they were the largest counties in different states, and the remainder were chosen by random sample from each region pool that qualified in the categories of population exceeding 500,000; between 100,000 and 500,000; and less than 100,000. An additional 139 locations were also included in the second stage to supplement the sample, which were selected based on targeted-sampling in districts that had experienced or retained targets likely to provoke terrorist activity. It is also mentioned that no sampling methodology was used in selecting state law enforcement and emergency agencies (Riley and Hoffman, 1995). The response rate for the study includes 39 state law enforcement agencies, 37 state emergency agencies, and 148 local law enforcement agencies (see Table 1 below). It is important to note that the study was begun in 1992 and completed in January 1993, one month before the 1994 World Trade Centre bombing. The significance of the timeliness of the study present evidence that the responses given in the survey were provided in a non-crisis state and therefore applicable to preparedness perceptions rather than lessons learned in hindsight. Any events of the terrorist bombing would therefore not be represented in the answers given.

Table 1. Response Rate of Research Sample

| Response Table<br>Agency Group | No. Agencies Invited | No. agencies Participated | Response Rate % |
|---|---|---|---|
| State Law Enforcement | 52 | 39 | 73% |
| State Emergency | 52 | 37 | 71% |
| Local Law Enforcement | 299 | 148 | 49% |
| Local Law Enforcement (population-based) | 160 | 84 | 53% |
| Local Law Enforcement (targeted-sample) | 139 | 64 | 46% |

The first stage of preparing the data required a thorough exploration of the survey instrument to identify possible questions that provided relational data to assess the respondent's social network, or questions relevant to an analysis of the current perceptions of their coordination abilities. In searching for networking data, two questions were found providing information on the respondent's perceived interaction with other agencies (Figure 3 and 4 below). These two questions were combined to form the respondent's ego-centric network, which was used to analyze the social networking measures of egobetweenness and the degree of agency interconnectedness, as well as the respondents' sub-group structure based on a clique analysis.

```
B9. Which agencies do you coordinate with during terrorist
      investigations? (Circle all that apply.)

    FBI       1. FBI.
    ENERGY    2. Department of Energy.
    FAA       3. FAA.
    STATED    4. Department of State.
    SS        5. United States Secret Service.
    TRANS     6. Department of Transportation.
    CUSTOMS   7. United States Customs Service.
    STLAW     8. State law enforcement agencies.
    STOTH     9. Other state agencies.
    INTER    10. International agencies.
    LOCTRAN  11. State or local transportation agencies.
    COUNTY   12. County or local law enforcement agencies.
    OTHER    13. Other, please specify
```

```
A20. Has your department participated in any JOINT TRAINING
     EXERCISES with: (Circle all that apply.)

JTFBI     1. FBI
JTSTAT    2. Other state agencies -- in state
JTSTOTH   3. Other state agencies -- out of state,
JTCOUNTY  4. County or municipal agencies
JTSS      5. U.S. Secret Service
JTDEA     6. DEA
JTBP      7. Border Patrol
JTENERGY  8. Department of Energy
JTPROF    9. Professional associations, fraternal organizations,
             informal working group or private agency;
             Names ______________________________
JTPRIV   10. Private businesses
JTOTH    11. Other, specify _____________________
JTOTCOD      ____________________________________
```

Further investigation of the survey instrument presented the final measure for network connectedness as tie strength by ranking the frequency of contact between the respondent's agency and others of a particular group such as municipal or state agencies (see Figure 5 below). These questions were combined to give a single representation of tie strength. Figure 6 below Relational Frequency of contact question representing tie strength, study by Riley and Hoffman (1995).

| A6. How often does your department meet or exchange information on terrorism with COUNTY or MUNICIPAL AGENCIES? | A7. How often does your department meet or exchange information on terrorism with other STATE AGENCIES? |
|---|---|
| MEETMUN 1. Once a week or more. | MEETSTA 1. Once a week or more. |
| 2. Two or three times month. | 2. Two or three times month. |
| 3. Once every month or two. | 3. Once every month or two. |
| 4. A few times a year. | 4. A few times a year. |
| 5. Annually. | 5. Annually. |
| 6. Never. | 6. Never. |

Exploration for coordination-based questions revealed a single item applicable for two independent measures of coordination (see figure 7). They are represented as accessibility of information, which is defined in this question by the number of sources used, and quality which is defined as a rank of usefulness from the sources used. The final measure of coordination, defined as readiness was extracted from a question which asked the respondent how prepared they perceive their agency to be to respond to an incident such as a terrorist disaster event (see figure 8 below).

D4. The following is a list of possible sources of information pertaining to terrorism. How useful have you found these sources to be?

| | | Never Used | Not Useful | Somewhat Useful | Very Useful |
|---|---|---|---|---|---|
| IFFBI | 1. FBI unclassified reports. | 1 | 2 | 3 | 4 |
| IFFBICS | 2. FBI classified reports. | 1 | 2 | 3 | 4 |
| IFFED | 3. Other federal agencies. | 1 | 2 | 3 | 4 |
| IFSTAT | 4. State agencies. | 1 | 2 | 3 | 4 |
| IFLOC | 5. Local jurisdictions. | 1 | 2 | 3 | 4 |
| IFMED | 6. The media. | 1 | 2 | 3 | 4 |
| IFPROF | 7. Professional law enforcement publications. | 1 | 2 | 3 | 4 |
| IFRISK | 8. Risk assessment services or publications | 1 | 2 | 3 | 4 |
| IFBOOK | 9. Books, journals, periodicals, non-law enforcement publications. | 1 | 2 | 3 | 4 |
| IFRAD | 10. Radical publications, other "alternative" literature. | 1 | 2 | 3 | 4 |
| IFINFORM | 11. Informants, sources on the street. | 1 | 2 | 3 | 4 |
| IFOTH | 12. Other ____________________ | 1 | 2 | 3 | 4 |

D10. How well prepared are you to respond to such an incident?

PREPARED
1. Very well prepared.
2. Well prepared.
3. Somewhat prepared.
4. Not well prepared.

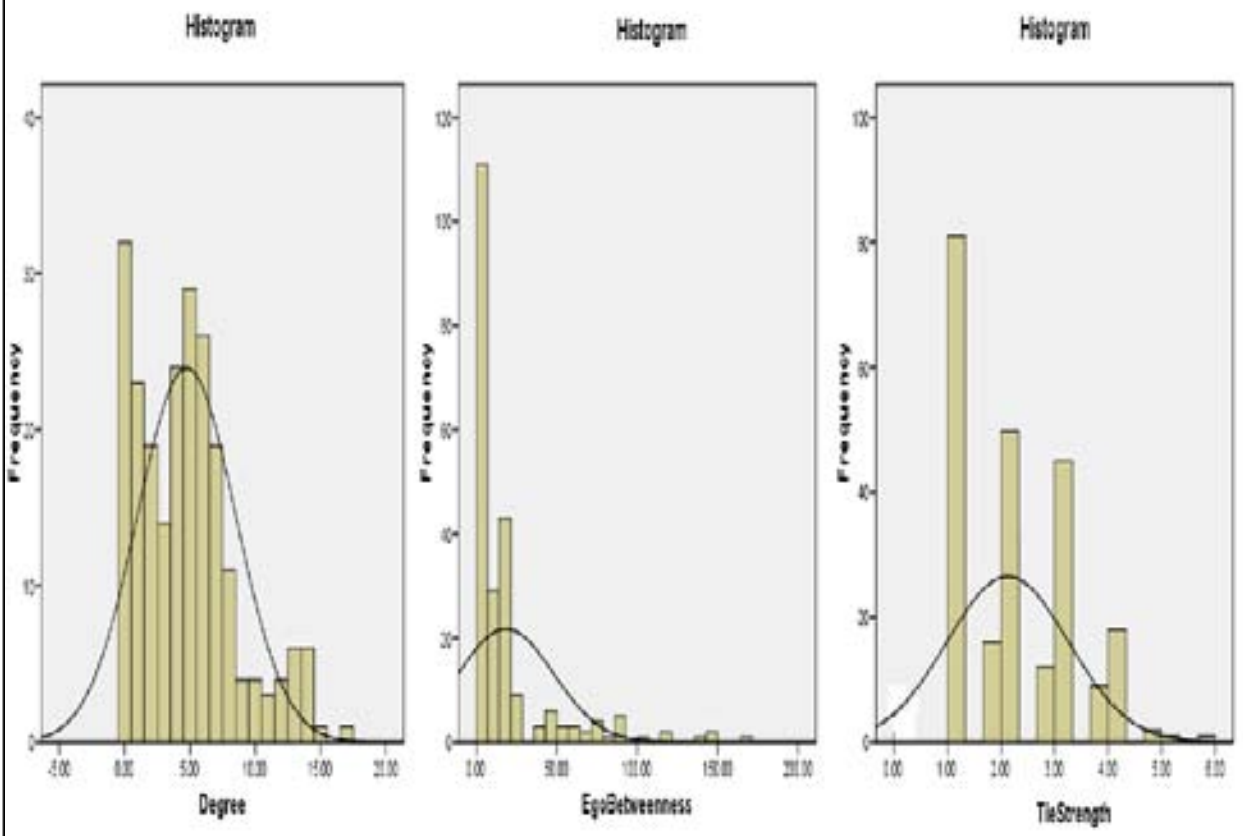

An examination of the Degree, EgoBetweeness and Tie Strength measures using SPSS reveals common distributions of all three that follow a non-normal curve. Each graph consists of centralized scores with a tapered skew to the right. This distribution is set against a line to illustrate the scores needed to represent a bell-curve. As a result of the non-normal distribution, non-parametric statistical testing must be carried out. The model uses the Kruskal-Wallis test to compare the mean ranks of the interconnectedness scores. This test is a non-parametric substitute for a one-way ANOVA comparison. The other test used in the model is correlations, which is examined by means of a Spearman Rank Order Correlation. The Spearman test is a non-parametric alternative to the Pearson test, which investigates the relationship between two continuous scores. We use the Spearman test rather than a Chi-square analysis because all of the variables being explored are created as continuous and as such require tests involving two or more continuous variables. By placing the variables into SPSS, we are able to perform some statistical analyses for hypothesis testing as defined in our DRN model (see Table 2 below).

Table 2. SPSS Test Matrix Defining Tests to Be Carried Out and Which Variables to Be Used

| **SPSS Test Matrix** | Degree | EgoBetweenness | Tie Strength | Quality | Accessibility | Readiness |
|---|---|---|---|---|---|---|
| Degree | Kruskal-Wallis Comparison | x | x | x | x | x |
| EgoBetweenness | x | Kruskal-Wallis Comparison | x | x | x | x |
| Tie Strength | x | x | Kruskal-Wallis Comparison | x | x | x |
| Quality | Spearman Correlation | Spearman Correlation | Spearman Correlation | x | x | x |
| Accessibility | Spearman Correlation | Spearman Correlation | Spearman Correlation | x | x | x |
| Readiness | Spearman Correlation | Spearman Correlation | Spearman Correlation | x | x | x |

Note: x denotes unnecessary or out of scope testing

A key point to note about the study by Riley and Hoffman (1992) is that the data collection method was by means of a survey instrument. Sinclair (1995) notes that survey's, along with any sort of questionnaire or interview methodology for gathering data is described as being subjective. Blyth (1972) defines subjective data as retaining personal beliefs and incorporating pre-judgments rather than simply providing impartial facts. The relational data collected for social networking analysis is a subjective perception of the respondent's emergency contacts as it requires the respondent to remember, circle and list all actors with whom he/she exchanges within the DRN. In contrast to this, the study by Naim Kapucu (2005) on the DRN during the September 11 Terrorist disaster is an account of the actual network as it existed during the crisis. Kapucu's (2005) study looks at the situation reports of the event as collected by FEMA. This data collection methodology retains objective measures as according to McClelland (1995), objective data may be a direct record registered by an independent observer in the form of video, audio or, in this case, text. The composition of the questions pertinent to coordination in Riley and Hoffman’s (1995) study are devised of rank and ratings scales, both of which are mentioned by Sinclair (1995) as common subjective data collection methods and therefore open to the same discrimination as the relational questions. Blyth (1972) states that the most important aspect to remember when studying subjective data is that although it is useful to analyze for a given perspective, a major disadvantage is that the beliefs or perception of the respondent may be wrong or only partially inclusive or accurate. It is assumed that such is the case with the relational questions being analyzed and that the egocentric network only contains partial information on alters in respondents’ network.

## RESULTS AND DISCUSSIONS

We first provide an overview of the high level organizational network from the data set. Second, we provide testing of our hypotheses at both a macro and micro level to explore interactions of actors and their organization in the context of responding to emergency situation. Lastly, we provide analysis which is statistically significant to validate and justify the relationship between coordination and social network by testing our DRN model and the hypotheses developed within. The high level representation (Figure 9) illustrates the organizations involved in DRN from an ego perspective.

Figure 9. Macro-level Representation of the State and Local Agency Ego Networks

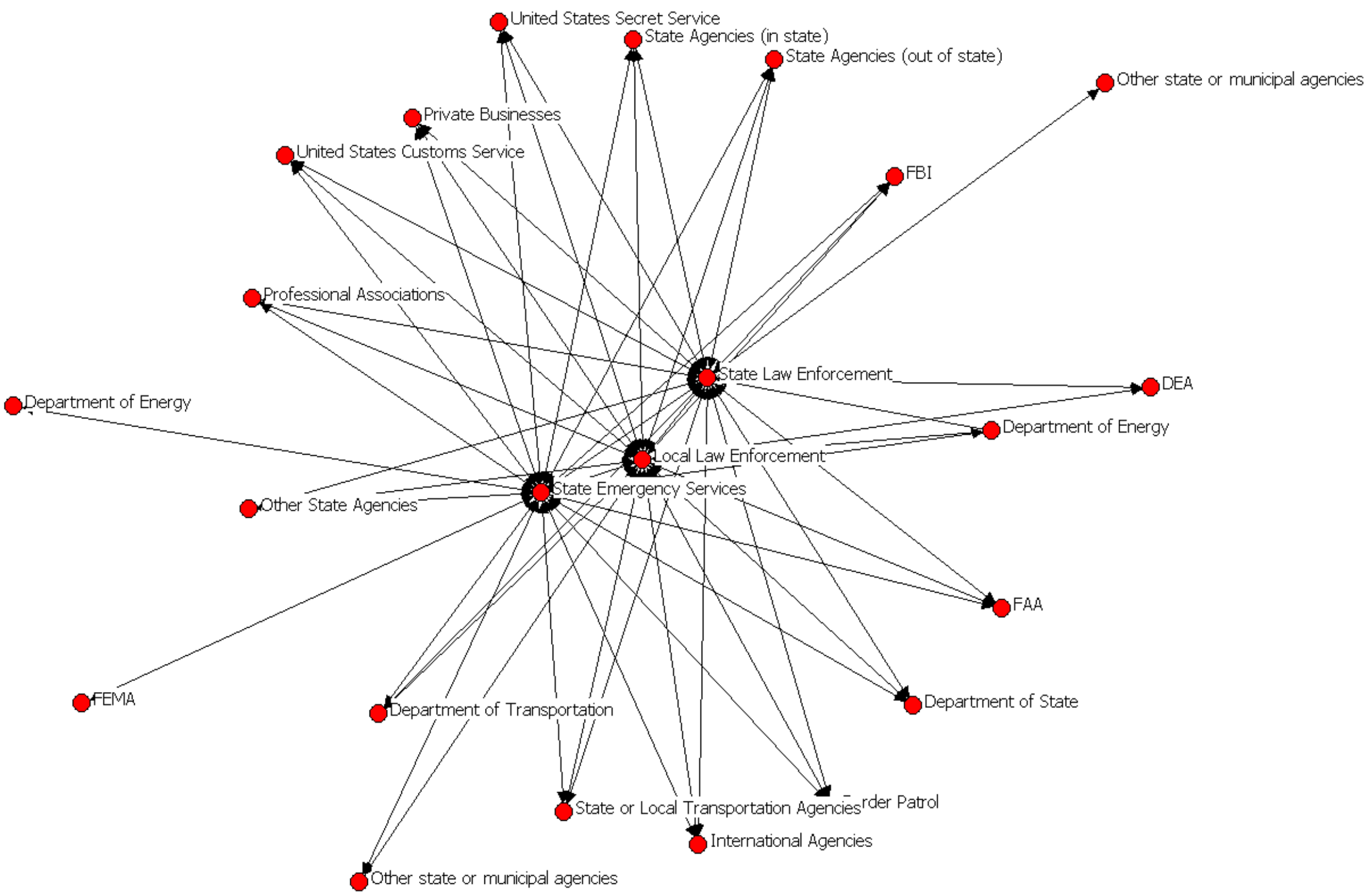

We first explore the network from a macro level to determine if any organizations naturally work together and form closer relationships. This clustering of agencies is supported through the identification and cooperation of individual actors representing their respective agencies at a micro-level. After examining inter-agency integration, the concept of tier mediation over network involvement is investigated. We first look at this notion at a macro-level by comparing agency networks, and then study this idea more closely at the micro-level by comparing clusters. A final examination of the relationship between network measures is assessed against coordination to determine if a correlation exists. A general hypothesis of correlation is investigated and therefore this is analyzed at the macro-level. To provide evidence that the same concept can be applied at any level of the network, the micro-level clusters are also examined for correlations between interconnectedness and coordination. The purpose of this section is to provide evidence to support the DRN to generate discussion and further investigation of the model with agencies from different business sectors in an effort to increase coordination preparedness by optimizing DRN conditions. The network visualization below (see Figure 10) illustrates the combined respondent data from all three agencies into one egocentric network. At first glance the network shows a similar level of centrality between all three organizational actors, which implies that all three organizations may have a similar level of network involvement.

Figure 10. Micro-level representation of the combined state law, state emergency and local law enforcement networks

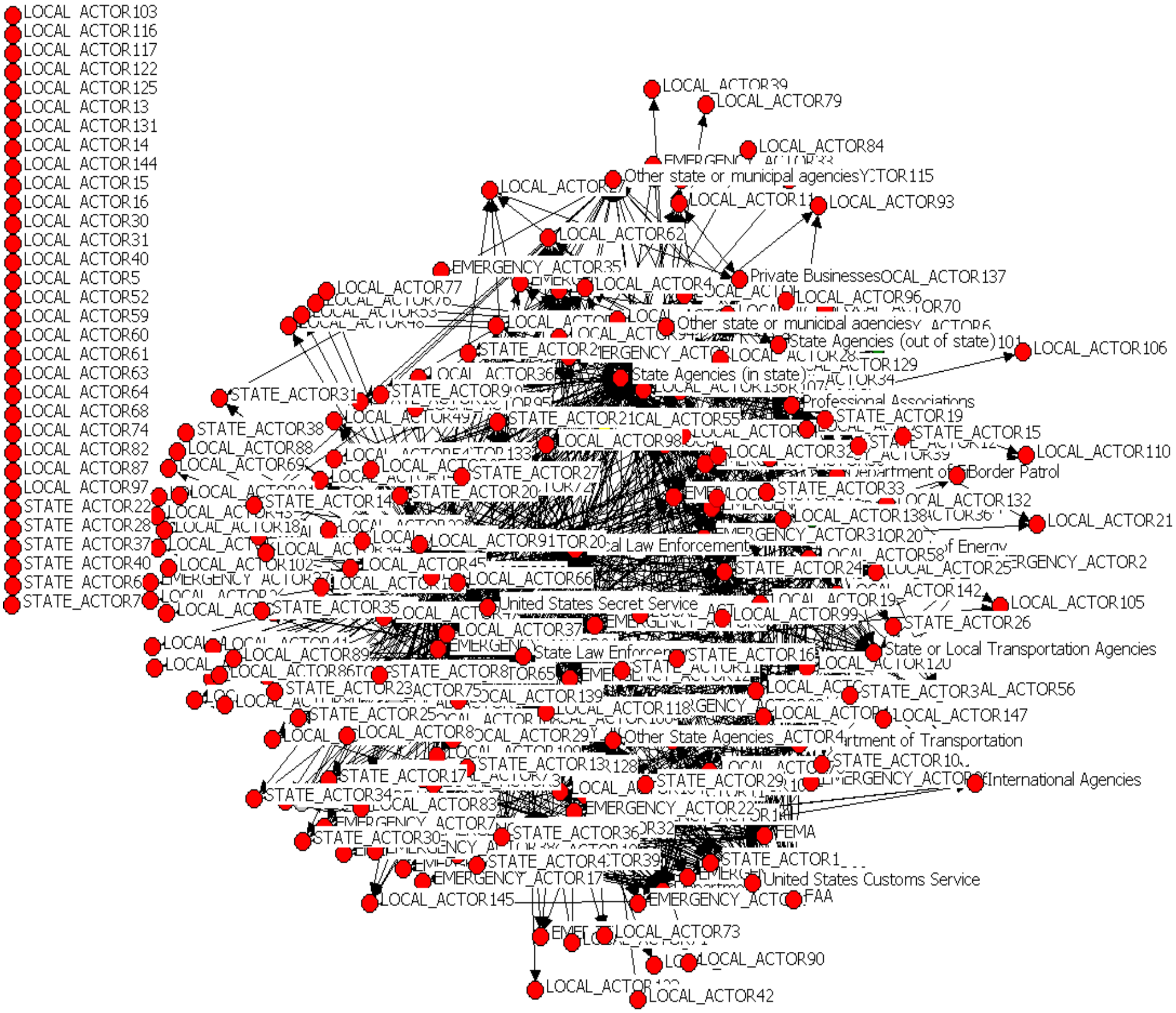

It has previously been discussed that although DRN follow a distributed approach to responding to extreme disaster events, there exists a need for organizational leadership based on a control or executive order structure, which is defined in the DRN model as tiers (Department of Homeland Security, 2006; Kettl, 2006). The National Response Plan (Department of Homeland Security, 2006) documents decision-making roles and responsibilities and control of the network based on this loose leadership hierarchy (see Figure 11, Department of Homeland Security, 2006 for evidence of this hierarchy). Although the NRP does not clearly state how many tiers this structure has, or a definitive flow of leadership, it does provide some evidence to suggest that once a “Presidential disaster or emergency declaration” has been made, the first-order coordination responsibilities falls to federal government agencies such as FEMA. The plan implies that government agencies other than at federal level share an interdependent role under federal leadership to then guide organizations from other sector. This structure, insinuated in the response plan frameworks developed by the Department of Homeland Security (Figure 11), leads to the theory of a 3-tiered control approach to emergency coordination within the distributed network.

Figure 11. Evidence of hierarchical leadership and coordination in emergency response, Department of Homeland Security, 2006

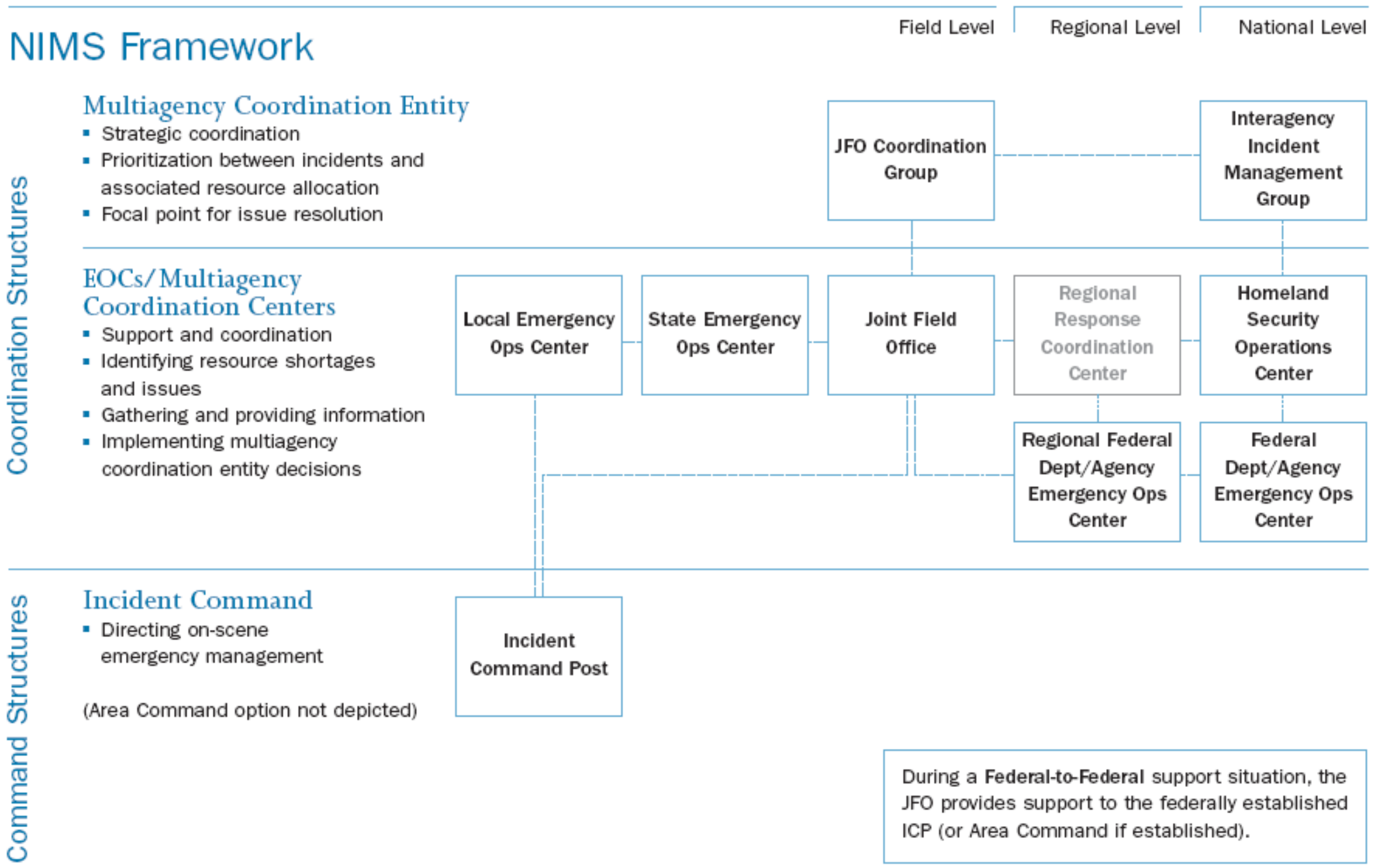

As evidence to suggest that local and state government agencies share the 2nd tier of network control, a hypothesis is put forward which states:

(H1) – *A prediction of what tier an organization belongs to can be made by analyzing an organization for its subgroup*.

*Organizational Clique Analysis*

A clique analysis of Figure 10 above reveals that there are 15 cliques (see Table 3 and Table 4 below) within the perceived ego network of the three groups, all of which contain state and local agencies, and some of which include federal agencies. This overlap with some federal agencies is natural in an emergency response network as organizations need to exchange information and resources to coordinate through the distributed structure. This analysis suggests that local and state agencies do in fact share a common level of involvement within the DRN For a more definitive confirmation of this finding, however, an analysis of agencies in other tiers would need to be tested.

Table 3. Clique analysis output of respondents from state and local agencies

```
15 cliques found.

   1:  Department of State Local Law Enforcement State Emergency Services State Law Enforcement
   2:  Department of Transportation Local Law Enforcement State Emergency Services State Law Enforcement
   3:  FAA Local Law Enforcement State Emergency Services State Law Enforcement
   4:  FBI Local Law Enforcement State Emergency Services State Law Enforcement
   5:  International Agencies Local Law Enforcement State Emergency Services State Law Enforcement
   6:  Border Patrol Local Law Enforcement State Emergency Services State Law Enforcement
   7:  Local Law Enforcement Other State Agencies State Emergency Services State Law Enforcement
   8:  Local Law Enforcement Private Businesses State Emergency Services State Law Enforcement
   9:  Local Law Enforcement Professional Associations State Emergency Services State Law Enforcement
  10:  Local Law Enforcement State Agencies (in state) State Emergency Services State Law Enforcement
  11:  Local Law Enforcement State Agencies (out of state) State Emergency Services State Law Enforcement
  12:  Local Law Enforcement State Emergency Services State Law Enforcement
  13:  Local Law Enforcement State Emergency Services State Law Enforcement State or Local Transportation Agencies
  14:  Local Law Enforcement State Emergency Services State Law Enforcement United States Customs Service
  15:  Local Law Enforcement State Emergency Services State Law Enforcement United States Secret Service
```

Table 4. Clique co-membership matrix to determine sub-groupings

Organisation-by-Organisation Clique Co-Membership Matrix

| | | 1 De | 2 Bo | 3 DE | 4 Fi | 5 De | 6 De | 7 FA | 8 FB | 9 FE | 10 In | 11 Lo | 12 Ot | 13 Ot | 14 Ot | 15 Pr | 16 Pr | 17 St | 18 St | 19 St | 20 St | 21 St | 22 Un | 23 Un |
|---|---|---|---|---|---|---|---|---|---|---|---|---|---|---|---|---|---|---|---|---|---|---|---|---|
| 1 | Department of Energy | 0 | 0 | 0 | 0 | 0 | 0 | 0 | 0 | 0 | 0 | 0 | 0 | 0 | 0 | 0 | 0 | 0 | 0 | 0 | 0 | 0 | 0 | 0 |
| 2 | Border Patrol | 0 | 1 | 0 | 0 | 0 | 0 | 0 | 0 | 0 | 0 | 1 | 0 | 0 | 0 | 0 | 0 | 0 | 0 | 1 | 1 | 0 | 0 | 0 |
| 3 | DEA | 0 | 0 | 0 | 0 | 0 | 0 | 0 | 0 | 0 | 0 | 0 | 0 | 0 | 0 | 0 | 0 | 0 | 0 | 0 | 0 | 0 | 0 | 0 |
| 4 | Department of Energy | 0 | 0 | 0 | 1 | 0 | 0 | 0 | 0 | 0 | 0 | 1 | 0 | 0 | 0 | 0 | 0 | 0 | 0 | 1 | 1 | 0 | 0 | 0 |
| 5 | Department of State | 0 | 0 | 0 | 0 | 1 | 0 | 0 | 0 | 0 | 0 | 1 | 0 | 0 | 0 | 0 | 0 | 0 | 0 | 1 | 1 | 0 | 0 | 0 |
| 6 | Department of Transportation | 0 | 0 | 0 | 0 | 0 | 1 | 0 | 0 | 0 | 0 | 1 | 0 | 0 | 0 | 0 | 0 | 0 | 0 | 1 | 1 | 0 | 0 | 0 |
| 7 | FAA | 0 | 0 | 0 | 0 | 0 | 0 | 1 | 0 | 0 | 0 | 1 | 0 | 0 | 0 | 0 | 0 | 0 | 0 | 1 | 1 | 0 | 0 | 0 |
| 8 | FBI | 0 | 0 | 0 | 0 | 0 | 0 | 0 | 1 | 0 | 0 | 1 | 0 | 0 | 0 | 0 | 0 | 0 | 0 | 1 | 1 | 0 | 0 | 0 |
| 9 | FEMA | 0 | 0 | 0 | 0 | 0 | 0 | 0 | 0 | 0 | 0 | 0 | 0 | 0 | 0 | 0 | 0 | 0 | 0 | 0 | 0 | 0 | 0 | 0 |
| 10 | International Agencies | 0 | 0 | 0 | 0 | 0 | 0 | 0 | 0 | 0 | 1 | 1 | 0 | 0 | 0 | 0 | 0 | 0 | 0 | 1 | 1 | 0 | 0 | 0 |
| 11 | Local Law Enforcement | 0 | 1 | 0 | 1 | 1 | 1 | 1 | 1 | 0 | 1 | 15 | 1 | 0 | 0 | 1 | 1 | 1 | 1 | 15 | 15 | 1 | 1 | 1 |
| 12 | Other State Agencies | 0 | 0 | 0 | 0 | 0 | 0 | 0 | 0 | 0 | 0 | 1 | 1 | 0 | 0 | 0 | 0 | 0 | 0 | 1 | 1 | 0 | 0 | 0 |
| 13 | Other state or municipal agencies | 0 | 0 | 0 | 0 | 0 | 0 | 0 | 0 | 0 | 0 | 0 | 0 | 0 | 0 | 0 | 0 | 0 | 0 | 0 | 0 | 0 | 0 | 0 |
| 14 | Other state or municipal agenciesY | 0 | 0 | 0 | 0 | 0 | 0 | 0 | 0 | 0 | 0 | 0 | 0 | 0 | 0 | 0 | 0 | 0 | 0 | 0 | 0 | 0 | 0 | 0 |
| 15 | Private Businesses | 0 | 0 | 0 | 0 | 0 | 0 | 0 | 0 | 0 | 0 | 1 | 0 | 0 | 0 | 1 | 0 | 0 | 0 | 1 | 1 | 0 | 0 | 0 |
| 16 | Professional Associations | 0 | 0 | 0 | 0 | 0 | 0 | 0 | 0 | 0 | 0 | 1 | 0 | 0 | 0 | 0 | 1 | 0 | 0 | 1 | 1 | 0 | 0 | 0 |
| 17 | State Agencies (in state) | 0 | 0 | 0 | 0 | 0 | 0 | 0 | 0 | 0 | 0 | 1 | 0 | 0 | 0 | 0 | 0 | 1 | 0 | 1 | 1 | 0 | 0 | 0 |
| 18 | State Agencies (out of state) | 0 | 0 | 0 | 0 | 0 | 0 | 0 | 0 | 0 | 0 | 1 | 0 | 0 | 0 | 0 | 0 | 0 | 1 | 1 | 1 | 0 | 0 | 0 |
| 19 | State Emergency Services | 0 | 1 | 0 | 1 | 1 | 1 | 1 | 1 | 0 | 1 | 15 | 1 | 0 | 0 | 1 | 1 | 1 | 1 | 15 | 15 | 1 | 1 | 1 |
| 20 | State Law Enforcement | 0 | 1 | 0 | 1 | 1 | 1 | 1 | 1 | 0 | 1 | 15 | 1 | 0 | 0 | 1 | 1 | 1 | 1 | 15 | 15 | 1 | 1 | 1 |
| 21 | State or Local Transportation Agencies | 0 | 0 | 0 | 0 | 0 | 0 | 0 | 0 | 0 | 0 | 1 | 0 | 0 | 0 | 0 | 0 | 0 | 0 | 1 | 1 | 1 | 0 | 0 |
| 22 | United States Customs Service | 0 | 0 | 0 | 0 | 0 | 0 | 0 | 0 | 0 | 0 | 1 | 0 | 0 | 0 | 0 | 0 | 0 | 0 | 1 | 1 | 0 | 1 | 0 |
| 23 | United States Secret Service | 0 | 0 | 0 | 0 | 0 | 0 | 0 | 0 | 0 | 0 | 1 | 0 | 0 | 0 | 0 | 0 | 0 | 0 | 1 | 1 | 0 | 0 | 1 |

By examining the combined state law, state emergency and local law enforcement networks at a micro level (see Figure 10), we can explore the clustering of individual actors to provide further evidence to support the first hypothesis that actors from within the three agency groups are interconnected and share responsibilities for driving the DRN.

*Organizational n-Clique Analysis (Micro level)*

Performing an n-clique analysis on the low-level network allows us to dissect the network and investigate network behaviour at a micro-level. This discrete perspective uncovers 249 clusters or sub-grouping within the greater network; closer inspection of the sub-groups reveals actors from each of the three agencies within a distance of length of 2 from each other. It is noted that there is an overlap with agencies between clusters, however for the purpose of this investigation, identification of clusters containing actors from both state and local agencies is sufficient evidence to support the hypothesis.

An assessment of three clusters selected at random from the n-clique analysis allows us to further investigate our DRN model and examine the behaviour of cross-sections of the DRN to support the stated hypotheses. The three diagrams of each of the clusters (see Figures 12, 13 and 14 below) provide a visual representation of the sub-groups under investigation. Note that within the sub-groups, only organizations from state law, state emergency and local law enforcement agencies are investigated. The graphs illustrate other agencies in pink that are identified to be within the clusters, however only the nodes in yellow, blue, and orange, respectively, will be assessed in the DRN model and compared. At face-value, it is evident that organizations from each of the three agencies work closely within the model as depicted in Figures 12, 13, and 14.

Figure 12. Illustrates the first randomly selected cluster from the n-clique analysis

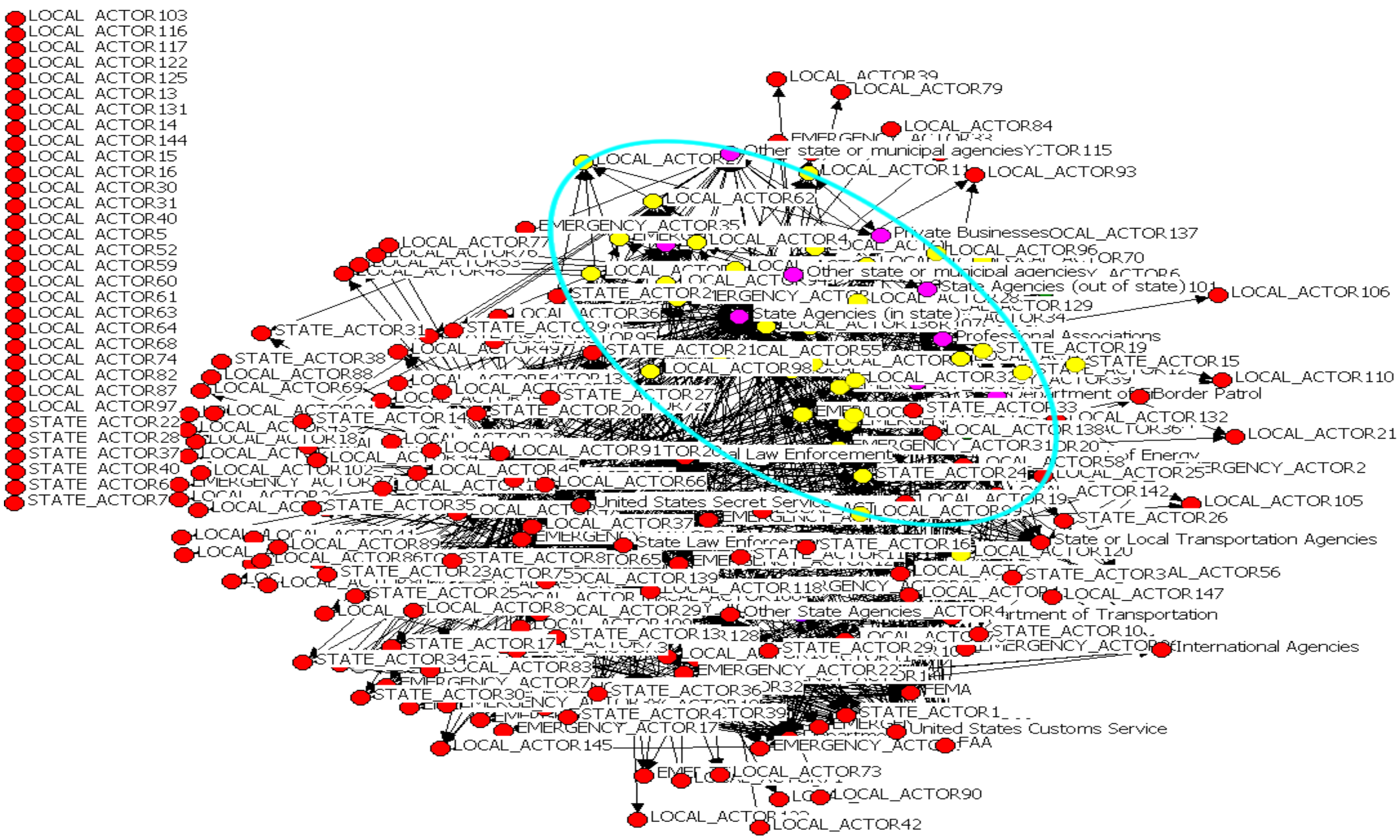

Figure 13. Illustrates the second randomly selected cluster from the n-clique analysis

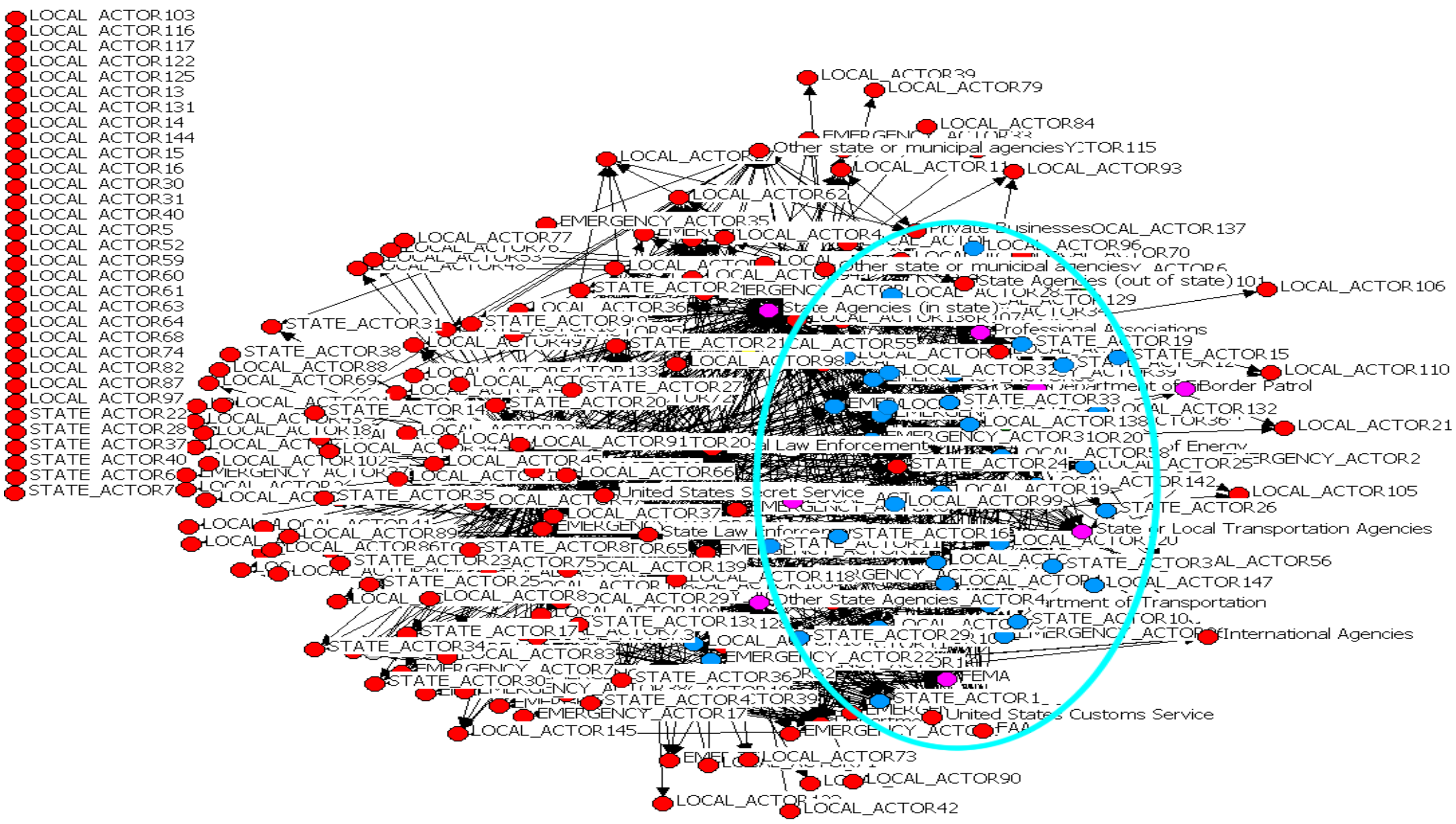

Figure 14. Illustrates the third randomly selected cluster from the n-clique analysis

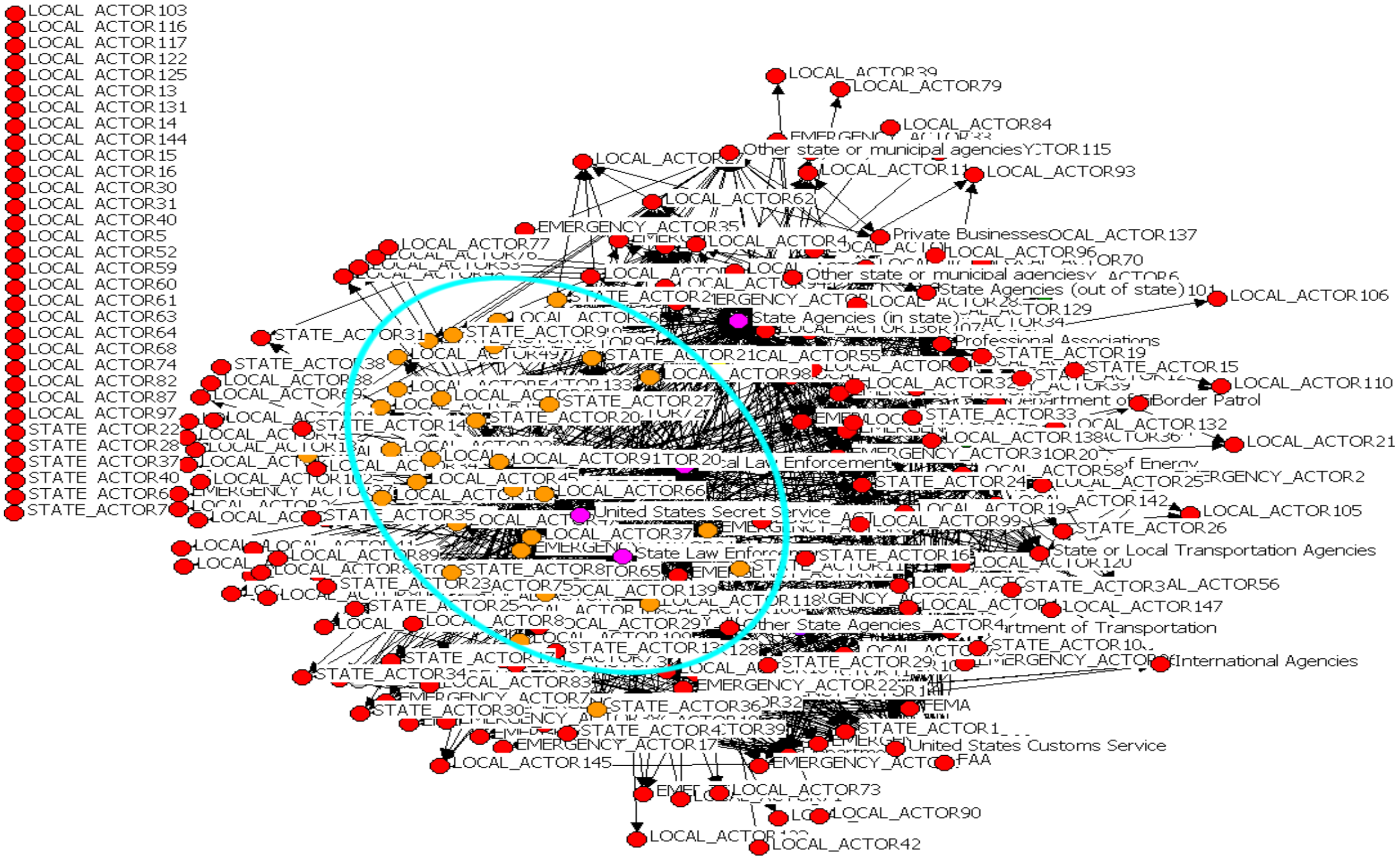

If we accept the test results above and assume that a clique analysis of an organization allows for tier definition, we can then test the second hypothesis that:

(H2*) The level of network involvement expected from an organization is mediated by the tier they fall into*.

Based on the clique results that state and local agencies share a common tier, we can perform a simple comparison to assess whether or not each second tier organizations share similar network connectedness as defined by an analysis of degree, EgoBetweenness and tie strength. The graphical data below illustrates the egocentric network of each of the three organizational actors' and how each individual respondent in that organization perceives their network (see Figures 15, 16, 17 below). Ultimately testing this hypothesis would need to be carried out with actual rather than perceived network data and involve studies of agencies from other tiers to decipher an accurate assessment of whether tier placement (which implies subgroup belonging) is a factor in how connected an organization is. At this point, however, data restrictions permit only a state and local assessment of the perceived network of the fore-mentioned government agencies.

Figure 15. State Law Enforcement Network

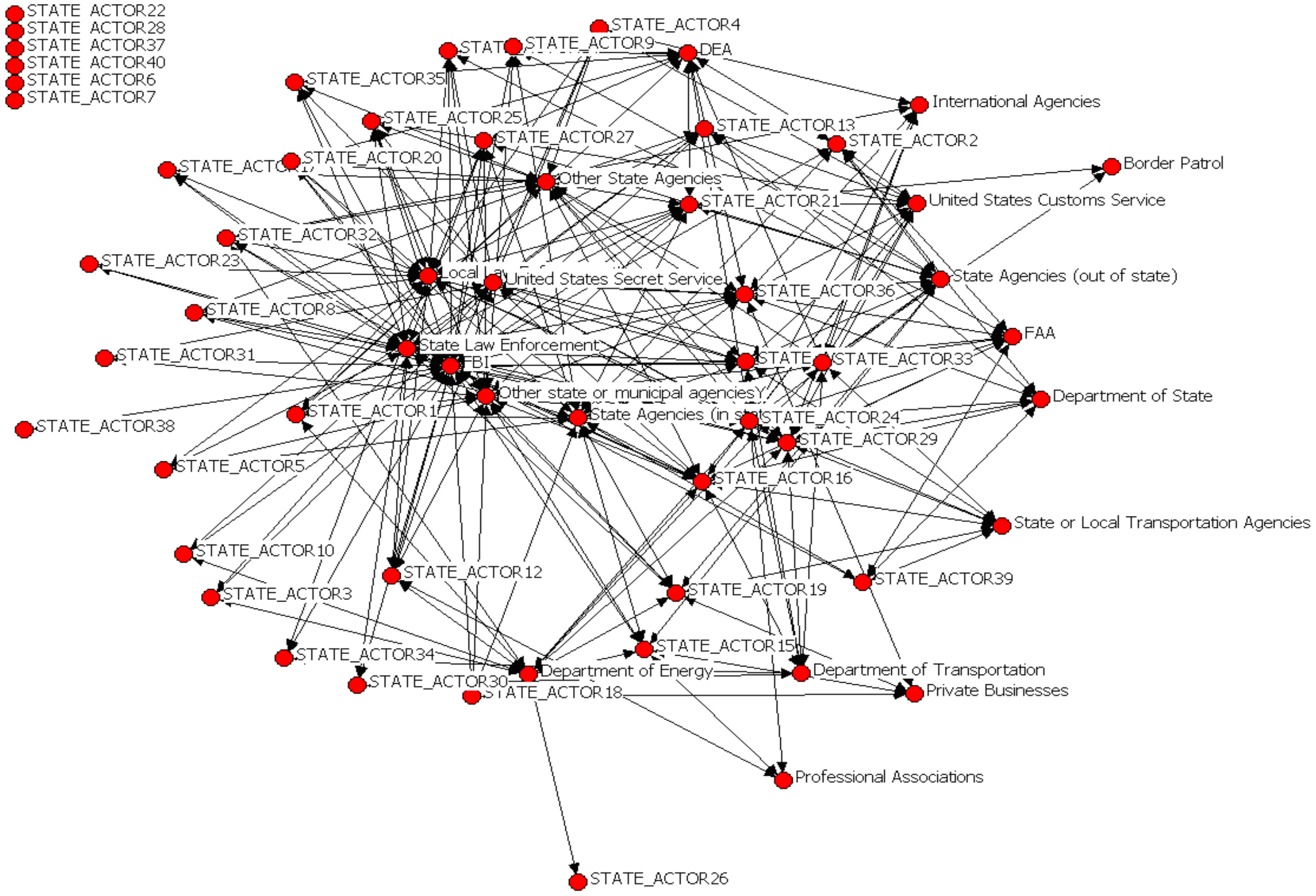

Figure 16. State Emergency Services network

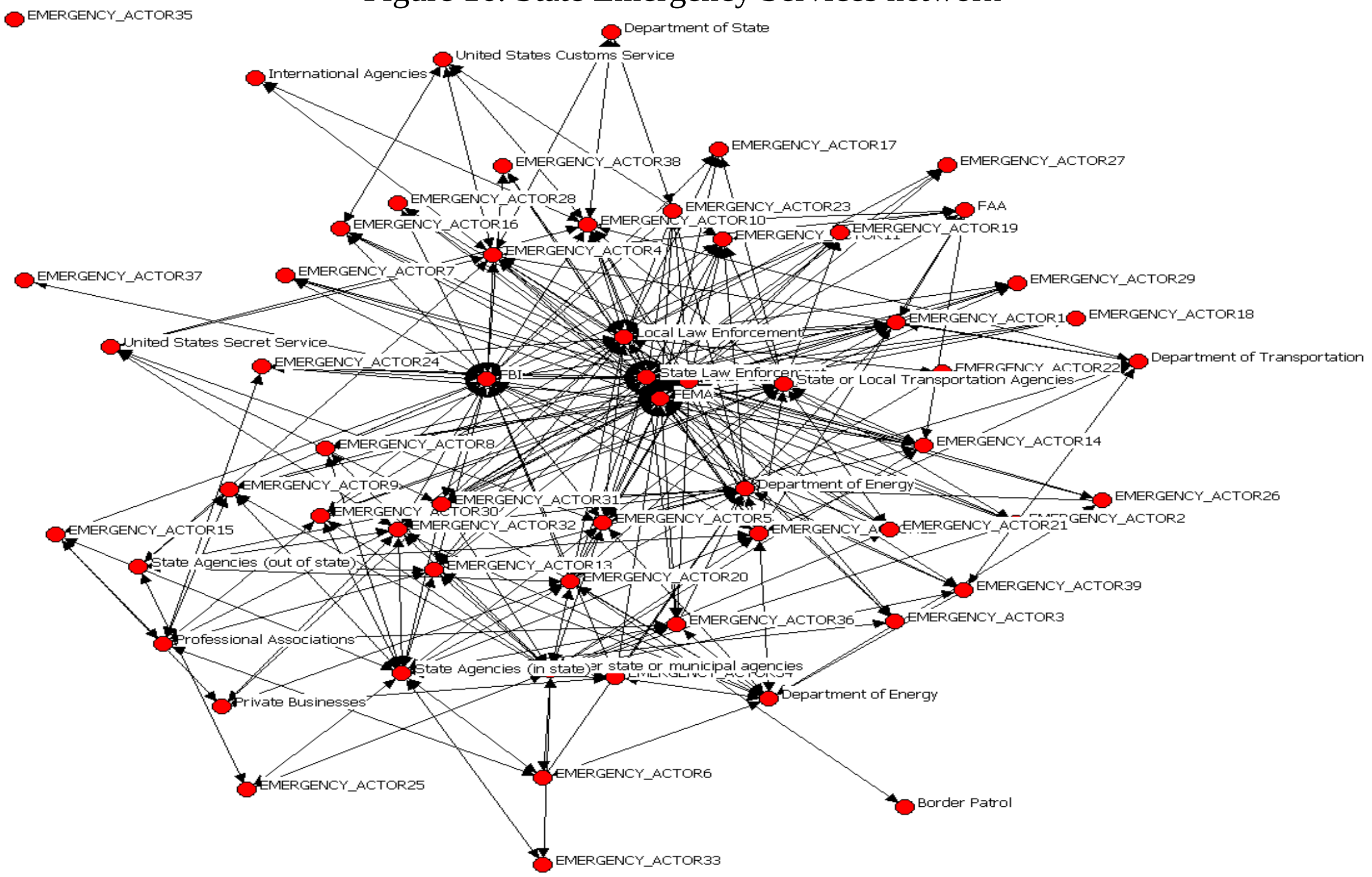

Figure 17. Local Law Enforcement network

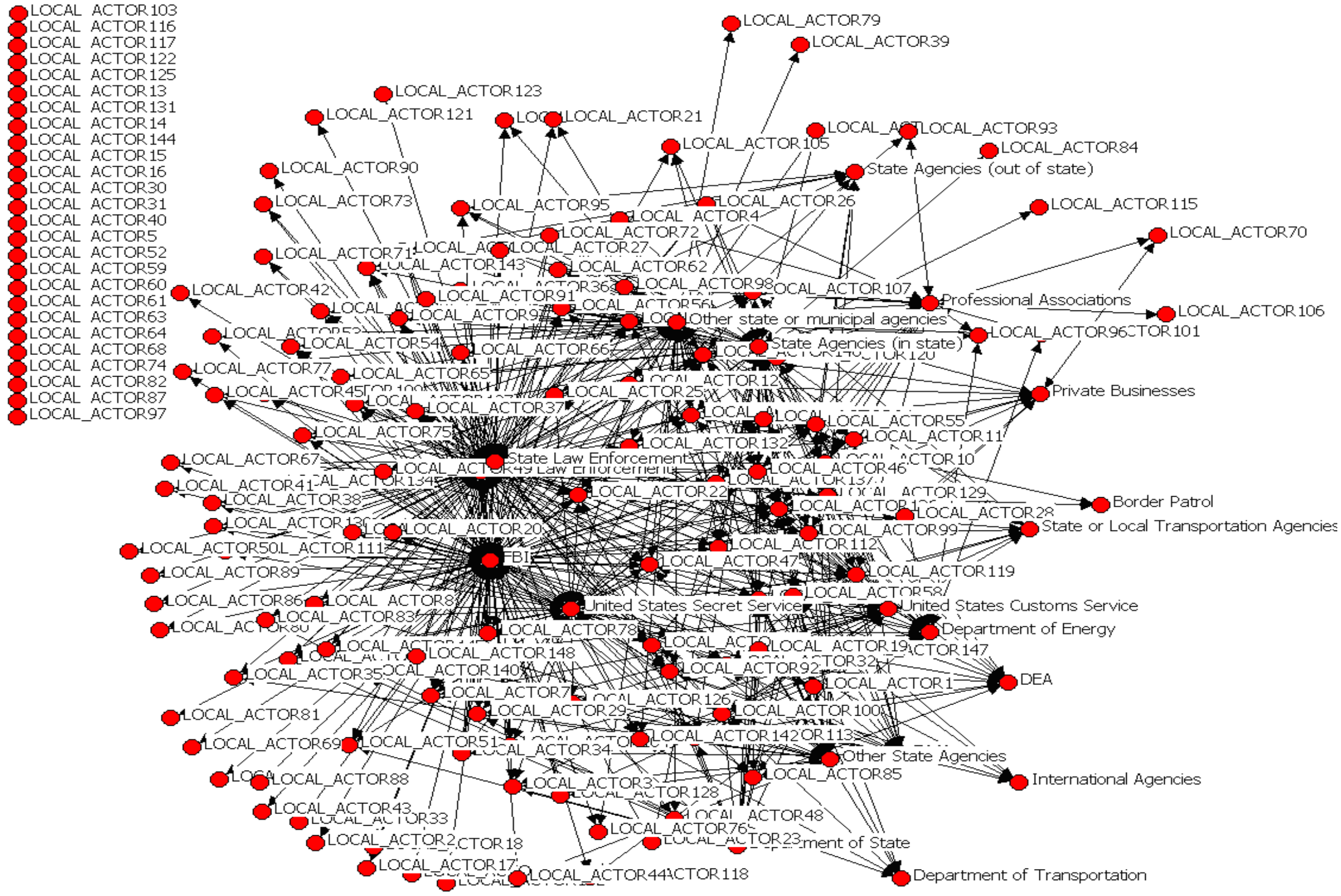

*Organizational comparison to determine tier placement and interconnectedness*

A Kruskal-Wallis test of the mean rank of each of the three interconnectedness measures shows a significant difference in scores between state law enforcement, state emergency services, and local law enforcement for degree and EgoBetweenness (Table 5 below). Mean rank scores of tie strength show no significant difference implying each agency maintains a similar range of weak and strong ties to organizations within the network.

Table 5. Kruskal-Wallis Interconnectedness Comparison of Mean Ranks

| *Kruskal-Wallis comparison of mean rank* | Degree | EgoBetweenness | Tie Strength |
|---|---|---|---|
| State Law Enforcement | 61.74 | 127.26 | 61.67 |
| State Emergency Services | 73.45 | 154.92 | 52.57 |
| Local Law Enforcement | 51.15 | 99.63 | 58.12 |
| Asymp. Sig. | 0.00** | 0.01** | 0.476 |

Note. Significant difference in rank mean at the 0.01 level are denoted **.
Significant difference in rank mean at the 0.05 level are denoted *.

By comparing the mean rank of state law enforcement agencies to state emergency services, a difference of 27.66 is produced under EgoBetweenness, and 11.71 for degree. This difference in scores is assumed to be an acceptable margin considering the NRP (Department of Homeland Security, 2006) identifies all state-based organizations as indifferent from each other and sharing common leadership responsibilities and connections within the DRN. When the mean rank of state and local law enforcement is compared for degree and EgoBetweenness, results indicate that local law enforcement is within the same margin to state law enforcement as state law is to state emergency. This suggests that although local law enforcement is somewhat less interconnected than state law and emergency services, it is still within the range of the two state agencies, and therefore possible to be includedin the same tier. The mean scores of tie strength produce no significant difference which suggests a similar network of weak and strong ties exist for each of the three agencies. This may be due to the tier structure as organizations within a specific subgroup would be likely to associate with the same agencies and seek information from similar contacts. Projection as to why there is a significant difference between scores of local law enforcement, state emergency services, and state law enforcement may be a consequence of task-based responsibilities. Each agency, although from the same tier still maintain specific functions unique to their organization within the network. For instance, local agencies such as law enforcement, fire and ambulance services perform hands-on tasks at ground zero, while state agencies provide more assistance and resource coordination. This requires an interdependency of government agencies at state and local level to provide leadership under the federal government for the DRN as a whole.

*Cluster comparison to determine tier placement and interconnectedness*

A Kruskal-Wallis test on each of the three clusters against the interconnectedness measures of Degree, EgoBetweenness and Tie Strength reveal no significant difference in scores (Table 6 below).

*Kruskal-Wallis comparison of mean rank*

| | Degree | EgoBetweenness | Tie Strength |
|---|---|---|---|
| Cluster 1 | 99.50 | 99.29 | 87.78 |
| Cluster 2 | 106.00 | 108.54 | 94.21 |
| Cluster 3 | 88.52 | 87.10 | 100.65 |
| Asymp. Sig. | 0.152 | 0.064 | 0.435 |

Note. Significant difference in rank mean at the 0.01 level are denoted **.
Significant difference in rank mean at the 0.05 level are denoted *.

The three clusters of state and local government agencies taken from the DRN, indicate that the level of connectedness between all three groups are comparable. The marginal differences in scores may suggest further evidence to the idea that although they are not significantly different, there is in fact a difference that may account for a threshold of network involvement allowing for an amount of variance between agencies or in this case clusters taken from the same tier. The fact that all three clusters, selected from what is believed to be the second tier, retain similar scores in the Kruskal-Wallis comparison is evidence suggesting that tier placement is a primary candidate for determining the level of network involvement. Although both of these tests are not definitive proof that an organization's interconnectedness is mediated by their tier level, it does in fact provide a step in the right direction by beginning to explore the possible thresholds within a given tier of what constitutes adequate network involvement and interconnectedness. It is only through an understanding of this threshold for each of the three tiers that an analysis can be carried out to clearly define whether interconnectedness is in fact mediated by the tier structure. Further analysis of clusters in different tier allocations would also be beneficial to support this hypothesis. Once an organizational analysis has been made to determine tier placement, it is proposed that a significant proportion of constitutes the effectiveness of coordination in DRN is the result of how involved an organization is within a network. This level of coordination based on network interconnection is arguably then a determining factor in what distinguishes an organization's state of preparedness. By using social networking theory to assess network involvement, the hypothesis put forward is that:

(H3) *There is a significant relationship between network involvement and coordination where an increase in network connectedness produces an increase in organizational coordination within a given threshold.*

This hypothesis is investigated by looking at the sub-hypotheses of social network measures correlating to measures of coordination in order to ascertain the overarching statement. The tests aim to provide evidence of a positive relationship between the two variables; literature, however, states that by being too involved in a network, too much information exchange can have a negative effect on coordination and efficiency (Kapucu, 2005; Scholten et al. 2005). This section of the hypothesis is for further investigation due to a lack of data for testing a threshold; however it is important to recognize the literature stating that a threshold exists, and excessive connectedness under the premise of network involvement as an enabler for coordination can be turned into an inhibitor.

*Connectedness correlates to increased coordination*

A Spearman test is used to determine if there is a relationship between the continuous independent connectedness variables of *degree, EgoBetweenness* and *tie strength* with the continuous dependent coordination variables of *readiness, quality* and *accessibility* (see Table 7 below). This test combines the agency data of all three networks to provide a general examination of the social networking measures against coordination measures.

Table 7. Spearman correlations matrix between connectedness and coordination

**Spearman correlations matrix (Combined agency network data)**

| | Degree | EgoBetweenness | Tie Strength | Readiness | Quality | Accessibility |
|---|---|---|---|---|---|---|
| Degree | 1 | | | | | |
| EgoBetweenness | x | 1 | | | | |
| Tie Strength | x | x | 1 | | | |
| Readiness | 0.263** | 0.252** | 0.221** | 1 | | |
| Quality | 0.231** | 0.216** | 0.281** | x | 1 | |
| Accessibility | 0.321** | 0.312** | 0.349** | x | x | 1 |

Note. Correlations significant at the 0.01 level (2-tailed) are denoted **.
Correlations significant at the 0.05 level (2-tailed) are denoted *.
'x' signifies correlations not tested

Increased Degree Centrality correlates to:

- (H3a) increased Coordination Readiness
- (H3b) increased Quality of Information
- (H3c) increased Information Accessibility

The results of the Spearman test indicate a positive correlation coefficient between degree and each of the three dependent coordination variables where an increase in the measure of degree produces an increase in readiness, information quality and accessibility; the tests all show significance at the 0.01 level. A positive increase to an organization’s readiness for an emergency and ability to be granted access to information that is also of a higher standard has shown to be, in part, a product of increasing the number of emergency contacts an organization maintains. This finding may arguably be a result of the nature of a distributed network structure. Kapucu (2005) states that not all organizations are central in DRN and can have effects on information transfer as it disseminates through the network. By increasing the number of contacts a given organization maintains, the network is able to become more connected and distributed as a whole, which can potentially enable better flow of information to reduce the coordination gap. From an individual organization's perspective this robustness created through increasing each organizational node's contacts producing a better connected network which may be the motivation for the improvement to aspects of coordination such as quality and accessibility of information, and how ready an organization is to respond in a crisis.

Increased EgoBetweenness correlates to:

- (H3d) increased Coordination Readiness
- (H3e) increased Quality of Information
- (H3f) increased Information Accessibility

The relationship between the three dependent coordination variables against EgoBetweenness produces a positive correlation coefficient to the 0.01 significance level. The Spearman test denotes an increase in each of the measures of coordination including readiness, quality and accessibility of information based on an increase in EgoBetweenness. The results of these tests provide evidence that an increase in an organizations ability to be in a controlling position within the DRN improves an organization's capacity to coordinate in an emergency. Potentially by being more central in the network and maintaining the capability to impede or enhance the flow of information, an organization may find itself in a more empowering position which dictates how accessible information is to come across, the quality of that information, and an overall readiness to coordinate with other nodes. Malone and Crowston (1994) state that the synchronization of information for coordination efficiency is of significant importance, by improving EgoBetweenness which has implications on network positioning and control (Freeman, 1979), an organization is able to be in a more dominant state to receive information and coordinate others.

Increased Tie Strength correlates to:

- (H3g) increased Coordination Readiness
- (H3h) increased Quality of Information
- (H3i) increased Information Accessibility

The Spearman correlation indicates that an increase in tie strength produces an increase in the quality of information, accessibility of information, and how ready an organization is to coordinate in an emergency; the results are significant to the 0.01 level.

An egocentric analysis of tie strength against coordination finds that an increase in the quality of relationships is able to improve coordination attributes such as quality and accessibility of information, and overall readiness for an emergency situation. Speculation as to why such a correlation exists may be due to the context of the data itself more than an overarching statement of tie strength. The study by Riley and Hoffman (1996) devises the question on tie strengths based on existing ties from local, state and federal departments. Under the framework of the original research study, it may be said that when organizations in an emergency network invest in existing relationships to strengthen the bond, interorganizational dependency becomes more efficient as trust is developed and collective sense making can be enhanced. This in turn may mean that after establishing better network relationships, an organization is more likely to have access to information that is of better quality due to other organizations being more forthcoming.

This improved working relationship may then be able to have positive affect on sharing which may facilitate coordination and perceived state of readiness to interact with other organizational nodes on an emergency.

*Connectedness correlates to increased coordination*

A subsequent examination for correlation is carried out to provide supporting evidence of the relationship between network connectedness and the potential for coordination. The three randomly selected clusters are merged to provide enough cases to perform a Spearman correlation and determine if the correlation that has been identified to exist at the macro-level of the DRN holds at the micro-level (see Table 8 below).

Spearman correlations matrix (combined cluster network data)

| | Degree | EgoBetweenness | Tie Strength | Readiness | Quality | Accessibility |
|---|---|---|---|---|---|---|
| Degree | 1 | | | | | |
| EgoBetweenness | x | 1 | | | | |
| Tie Strength | x | x | 1 | | | |
| Readiness | 0.112 | 0.100 | 0.220 | 1 | | |
| Quality | 0.292** | 0.264** | 0.219** | x | 1 | |
| Accessibility | 0.226** | 0.172* | 0.385** | x | x | 1 |

Note. Correlations significant at the 0.01 level (2-tailed) are denoted **.
Correlations significant at the 0.05 level (2-tailed) are denoted *.
'x' signifies correlations not tested

As discovered in the previous test, the cluster examination reveals a positive correlation between the coordination measures of *quality* and *accessibility* against the network interconnectedness measures of *degree*, *EgoBetweenness* and *tie strength*. This evidence supports the hypotheses that:

- (H3b) Increased *Degree* correlates to increased Information *Accessibility*
- (H3c) Increased *Degree* correlates to increased *Quality* of Information
- (H3e) Increased *EgoBetweenness* correlates to increased Information *Accessibility*
- (H3f) Increased *EgoBetweenness* correlates to increased *Quality* of Information
- (H3h) Increased *Tie Strength* correlates to increased Information *Accessibility*
- (H3i) Increased *Tie Strength* correlates to increased *Quality* of Information

The results for coordination readiness investigated at the micro-level reveal a difference in findings from the macro-level test. The data suggests a positive correlation between each of the three network connectedness measures to coordination readiness, however none of which were particularly significant. Speculation as to why this may have occurred may be a result of the clusters selected for analysis. Since there is evidence to suggest the interconnectedness scores of each cluster is not significantly different from each other as discovered while testing the second hypothesis, it is plausible that because of the macro-level correlation between interconnectedness and coordination readiness, all three clusters each provided similar subjective scores of readiness which as a result provided an insufficient range of readiness scores to calculate a correlation. To support this possible theory, a Kruskal-Wallis test is performed to determine if there is a significant difference between clusters. The results indicate that each of the three clusters are in very close proximity and therefore supports why no correlation was found between coordination readiness and interconnectedness at the micro-level (Table 9).

Table 9. Kruskal-Wallis comparison of readiness between clusters

*Kruskal-Wallis Comparison of Readiness*

| | Readiness |
|---|---|
| Cluster 1 | 99.52 |
| Cluster 2 | 95.70 |
| Cluster 3 | 94.64 |
| Asymp. Sig. | 0.868 |

Note. Significant difference in rank mean at the 0.01 level are denoted **.
Significant difference in rank mean at the 0.05 level are denoted *.

The work by Kapucu (2005) regarding interorganizational connectedness taken from the situation reports of the September 11 disaster is important for understanding the coordination gap that exists between the current state of ERNs and where the NRP (Department of Homeland Security, 2006) requires the standard of an emergency response to be. Arguably, the more organized and coordinated an emergency network is to respond to extreme events, the more likely a society is to have greater resilience to any form of disaster. Comments by Scholten et al. (2005), Stanley (2006) and Kettl (2006) mention lessons learned in hindsight of recent natural disasters such as Hurricane Katrina, and man-made disasters like the 2001 World Trade Center attacks focus on the need for better communication and coordination. It is stated that by having better coordination, the network can facilitate a more fluent exchange of information to enhance interorganizational collaboration (Scholten et al, 2005; Stanley, 2006; Kettl, 2006). Kapucu's (2005) exploratory study of ERN in crisis events highlights a significant lack of network connectedness when the emergency network is called on for a real-life response effort. Kapucu's (2005) evidence of structural holes in the network and weak points of communication, coupled with Scholten et al. (2005), Stanley (2006), and Kettl's (2006) account of the problems of an emergency response shows a coordination issue most probably brought about by a lack of network connectedness.

**CONCLUSION**

By presenting a model of coordination assessment based on network connectedness, an organization can be reviewed in order to find their current state of connectedness and therefore be judged for their potential to coordinate in an emergency. The findings from the hypothesis that (H1) *a prediction of what tier an organization belongs to can be made by analyzing an organization for their subgroup*, suggests that within DRN, organizations that share common traits form subgroups. These groupings then form the basis for an assumption that collective involvement from organizations that interact and share a common purpose within the DRN can be categorized into tiers which retain certain levels of authority and control over the network and the potential to coordinate in an emergency. The importance of this step in the DRN model is in making sure network connectedness acts as an enabler of coordination efficiency rather than an inhibitor by limiting network involvement to the needs of a given tier and thus preventing the circulation of redundant or unnecessary information through the network as a product of excessive ties. The influence for using a clique analysis for assessing organizations into tiers is based on the literature by Falzon (2000) where it is stated that, "*in any human organization in which individuals interact…groups emerge quite naturally and often deliberately…it helps us understand how information spreads throughout the organization*."

Although the data used for performing a clique analysis of an organization into tiers is somewhat limited due to the fixed list structure of the relational data questions in the study by Riley and Hoffman (1996), and by only examining organizations from the same tier, the test itself was able to give insight on how the theory would be carried out on organizations from all different levels. The results from this test provide evidence of the usefulness of tier assignment by means of cliques; validation of this hypothesis would, however, need to be carried out on a wider emergency audience.

The same data limitations present for the first hypothesis apply to the second hypothesis in that (H2) *the level of network involvement expected from an organization is mediated by the tier they fall into*. The results of the first hypothesis provide reasonable analysis to speculate that the organizations examined are derived from the same clique, and therefore represent a single tier in the DRN. The findings of the second hypothesis suggest a range of interconnectedness scores between these agencies which leads to the assumption that a threshold exists between organizations within the same tier. If this were in fact the case then it is reasonable to assume that since these agencies all retain network involvement scores within what is defined as the threshold of the second tier, then it is possible that tier allocation is a determining factor to the level of network involvement for a given organization.

This threshold may provide a rough guide for organizations based on their tier assessment level of how connected they should be within the DRN and of how much leadership they should show. The usefulness of the test results may account for the spectrum of network control between which nodes in the second tier currently operate. However to validate this statement, further analysis would need to be carried out on organizations in other tiers in order to provide a suitable comparison.

The hypothesis that (H3) *connectedness correlates to increased coordination* is important for understanding how a coordination gap can be minimized by establishing a greater connected network. The findings suggest that in order to investigate coordination in a distributed network, it is important to look beyond the task (Mintzberg, 1979) or relationship alone (Coordination Theory Model, Malone and Crowston, 1994) and examine the network structure and its implications on a coordination outcome. The analysis provides evidence of this by presenting a significant correlation between the measures of network involvement and the ability of that network to coordinate amongst themselves. The usefulness of these results show how an organization's perceived ability to coordinate is partially based on the fluency of the network itself. By combining these hypotheses, the model as a whole is able to assess how prepared an organization is to coordinate in an emergency based on how connected they are. Table 10 below illustrates each hypothesis's testing and the implications of each outcome on the model in order for coordination to occur within an ERN.

Table 10. Coordination Implications of Hypothesis Testing

| *Hypothesis Testing and Coordination Implications* | |
|---|---|
| Test | Implications of finding |
| T1a/b clique analysis of ERN actors | An actor's subgroup represents tier placement. |
| T2a/b Kruskal-Wallis SNA comparison | Actors from a single tier assert SNA scores within a threshold. |
| T3a/b Spearman correlation of connectedness to coordination | Network structure affects coordination Efficiency. |

The first hypothesis that subgroups help to delegate organizations into tiers is important to create horizontal and vertical network awareness and belonging to help organizations understand their role in an emergency response effort. The second hypothesis that an organization's tier moderates their connectedness is a necessary step for making sure that it is not simply the creation of a more connected network, but of a more efficiently connected network, where ties only exist where they are needed and to optimize information flow through the right channels. The third hypothesis that connectedness correlates to coordination establishes that by making a network more connected, coordination is enabled to be enhanced.